\documentclass[a4paper,fleqn,11pt]{article}

\usepackage{amsmath}
\usepackage{latexsym}
\usepackage{amssymb}
\usepackage{graphicx}
\usepackage{amsthm}

\newtheorem{thm}{Theorem}[section]
\newtheorem{lem}[thm]{Lemma}
\newtheorem{cor}[thm]{Corollary}

\theoremstyle{definition}
\newtheorem{defn}{Definition}[section]

\begin{document}
\title{Constructing Two Edge-Disjoint Hamiltonian Cycles and Two Equal Node-Disjoint Cycles in Twisted Cubes}
\author{\vspace{0.5cm}Ruo-Wei Hung\\
Department of Computer Science and Information Engineering,\\
Chaoyang University of Technology,\\
Wufong, Taichung 41349, Taiwan}

\maketitle

\begin{abstract}
The hypercube is one of the most popular interconnection networks
since it has simple structure and is easy to implement. The
$n$-dimensional twisted cube, denoted by $TQ_n$, an important
variation of the hypercube, possesses some properties superior to
the hypercube. Recently, some interesting properties of $TQ_n$
were investigated. In this paper, we construct two edge-disjoint
Hamiltonian cycles in $TQ_n$ for any odd integer $n\geqslant 5$.
The presence of two edge-disjoint Hamiltonian cycles provides an
advantage when implementing two algorithms that require a ring
structure by allowing message traffic to be spread evenly across
the twisted cube. Furthermore, we construct two equal
node-disjoint cycles in $TQ_n$ for any odd integer $n\geqslant 3$,
in which these two cycles contain the same number of nodes and
every node appears in one cycle exactly once. In other words, we
decompose a twisted cube into two components with the same size
such that each component contains a Hamiltonian cycle.\\

\noindent\textbf{Keywords:} edge-disjoint Hamiltonian cycles,
equal node-disjoint cycles, twisted cubes

\end{abstract}

\section{Introduction}
Parallel computing is important for speeding up computation. The
design of an interconnection network is the first thing to be
considered. Many topologies have been proposed in the literature
\cite{Bhuyan84,Choudum02,Cull95,Du88,Efe92,Hilbers87,Hwang00}, and
the desirable properties of an interconnection network include
symmetry, relatively small degree, small diameter, embedding
capabilities, scalability, robustness, and efficient routing.
Among those proposed interconnection networks, the hypercube is a
popular interconnection network with many attractive properties
such as regularity, symmetry, small diameter, strong connectivity,
recursive construction, partition ability, and relatively low link
complexity \cite{Saad88}. The architecture of an interconnection
network is usually represented by a graph. We will use graphs and
networks interchangeably.

The $n$-dimensional twisted cube $TQ_n$, an important variation of
the hypercube, was first proposed by Hilbers et al.
\cite{Hilbers87} and possesses some properties superior to the
hypercube. In fact, the twisted cube is derived from the hypercube
by twisting some edges. Due to these twisted edges, the diameter,
wide diameter, and fault diameter of $TQ_n$ are about half of
those of the comparable hypercube \cite{Chang99}. An
$n$-dimensional twisted cube is $(n-3)$-Hamiltonian connected
\cite{Huang02} and $(n-2)$-pancyclic \cite{Li06}, whereas the
hypercube is not. Moreover, its performance is superior to that of
the hypercube even if it is asymmetric \cite{Abraham91}.
Recently, some interesting properties of the twisted cube $TQ_n$
were investigated. Let $G$ be a graph. We denote by $V(G)$ and
$E(G)$ the node set and the edge set of $G$, respectively. A graph
$G$ is \textit{pancyclic} if, for every $4 \leqslant l \leqslant
|V(G)|$, $G$ has a cycle of length $l$. A graph $G$ is
\textit{edge-pancyclic} (resp. \textit{node-pancyclic}) if, for
any edge $e$ (resp. node $u$) of $G$ and every $4 \leqslant
l\leqslant |V(G)|$, $G$ has a cycle of length $l$ containing $e$
(resp. $u$). Yang et al. showed that, with $n_e+n_v\leqslant n-2$,
a faulty $TQ_n$ still contains a cycle of length $l$ for every
$4\leqslant l\leqslant |V(TQ_n)|-n_v$, where $n_e$ and $n_v$ are
the numbers of faulty edges and faulty nodes in $TQ_n$,
respectively \cite{Yang06}. In \cite{Fu08}, Fu showed that $TQ_n$
can tolerate up to $2n-5$ edge faults, while retaining a
fault-free Hamiltonian cycle. Fan et al. showed that the twisted
cube $TQ_n$, with $n\geqslant 3$, is edge-pancyclic and provided
an $O(l\log l+n^2+nl)$-time algorithm to find a cycle of length
$l$ containing a given edge of the twisted cube \cite{Fan08}. In
\cite{Fan08}, the author also asked if $TQ_n$ is edge-pancyclic
with $(n-3)$ faults for $n\geqslant 3$. Yang answered the question
and showed that $TQ_n$ is not edge-pancyclic with only one faulty
edge for any $n \geqslant 3$, and that $TQ_n$ is node-pancyclic
with $(\lfloor\frac{n}{2}\rfloor-1)$ faulty edges for every
$n\geqslant 3$ \cite{Yang09}. In addition, Lai et al. embedded a
family of 2-dimensional meshes into a twisted cube \cite{Lai08}.

Two Hamiltonian cycles in a graph are said to be
\textit{edge-disjoint} if they do not share any common edge. The
edge-disjoint Hamiltonian cycles can provide advantage for
algorithms that make use of a ring structure \cite{Rowley91}. The
following application about edge-disjoint Hamiltonian cycles can
be found in \cite{Rowley91}. Consider the problem of all-to-all
broadcasting in which each node sends an identical message to all
other nodes in the network. There is a simple solution for the
problem using an $n$-node ring that requires $n-1$ steps, i.e., at
each step, every node receives a new message from its ring
predecessor and passes the previous message to its ring successor.
If the network admits edge-disjoint rings, then messages can be
divided and the parts broadcast along different rings without any
edge contention. If the network can be decomposed into
edge-disjoint Hamiltonian cycles, then the message traffic will be
evenly distributed across all communication links. Edge-disjoint
Hamiltonian cycles also form the basis of an efficient all-to-all
broadcasting algorithm for networks that employ warmhole or
cut-through routing \cite{Lee90}.

Recently, conditional link faults of interconnected networks were
discussed in \cite{Fu08,Hsieh10}. The edge-disjoint Hamiltonian
cycles in $k$-ary $n$-cubes and hypercubes has been constructed in
\cite{Bae03}. Barden et al. constructed the maximum number of
edge-disjoint spanning trees in a hypercube \cite{Barden99}.
Petrovic et al. characterized the number of edge-disjoint
Hamiltonian cycles in hyper-tournaments \cite{Petrovic06}. Hsieh
et al. constructed edge-disjoint spanning trees in locally twisted
cubes \cite{Hsieh09}. Although the existence of a Hamiltonian
cycle in twisted cubes has been shown \cite{Huang02}, it is not
clear to generate edge-disjoint Hamiltonian cycles in twisted
cubes. In this paper, we show that, for any odd integer
$n\geqslant 5$, there are two edge-disjoint Hamiltonian cycles in
the $n$-dimensional twisted cube $TQ_n$. Further, we also
construct two equal node-disjoint cycles in the twisted cube. Two
cycles in a graph are said to be \textit{equal} and
\textit{node-disjoint} if they contain the same number of nodes,
there is no common node in them, and every node is in one cycle
exactly once. Finding two equal node-disjoint cycles in a
interconnected network is equivalent to decompose the network into
two disjoint components with the same number of nodes such that
each component contains a Hamiltonian cycle. Then, two distinct
algorithms that require a ring structure can be preformed in the
two components simultaneously. In this paper, we show that, for
any odd integer $n\geqslant 3$, there are two equal node-disjoint
cycles in the $n$-dimensional twisted cube $TQ_n$.

The rest of the paper is organized as follows. In Section
\ref{Preliminaries}, the structure of the twisted cube is
introduced, and some definitions and notations used throughout
this paper are given. Section \ref{EdgeDisjointHC} shows the
construction of two edge-disjoint Hamiltonian cycles in the
twisted cube. In Section \ref{NodeDisjointCyles}, we construct two
equal node-disjoint cycles in the twisted cube. Finally, we
conclude this paper in Section \ref{Conclusion}.

\section{Preliminaries}\label{Preliminaries}
We usually use a graph to represent the topology of an
interconnection network. A graph $G = (V, E)$ is a pair of the
node set $V$ and the edge set $E$, where $V$ is a finite set and
$E$ is a subset of $\{(u,v)| (u,v)$ is an unordered pair of $V\}$.
We also use $V(G)$ and $E(G)$ to denote the node set and the edge
set of $G$, respectively. If $(u,v)$ is an edge in a graph $G$, we
say that $u$ \textit{is adjacent to} $v$. A \textit{neighbor} of a
node $v$ in a graph $G$ is any node that is adjacent to $v$.
Moreover, we use $N_G(v)$ to denote the neighbors of $v$ in $G$.
The subscript `$G$' of $N_G(v)$ can be removed from the notation
if it has no ambiguity.

A path $P$, represented by $\langle v_0\rightarrow  v_1\rightarrow
\cdots\rightarrow v_{t-1} \rangle$, is a sequence of distinct
nodes such that two consecutive nodes are adjacent. The first node
$v_0$ and the last node $v_{t-1}$ visited by $P$ are called the
\textit{path-start} and \textit{path-end} of $P$, denoted by
$start(P)$ and $end(P)$, respectively, and they are called the
\textit{end nodes} of $P$. Path $\langle v_{t-1}\rightarrow \cdots
\rightarrow v_1 \rightarrow v_0 \rangle$ is called the
\textit{reversed path}, denoted by $P_{\textrm{rev}}$, of $P$.
That is, $P_{\textrm{rev}}$ visits the vertices of $P$ from
$end(P)$ to $start(P)$ sequently. In addition, $P$ is a cycle if
$|V(P)|\geqslant 3$ and $end(P)$ is adjacent to $start(P)$. A path
$\langle v_0\rightarrow v_1\rightarrow \cdots\rightarrow
v_{t-1}\rangle$ may contain other subpath $Q$, denoted as $\langle
v_0\rightarrow v_1\rightarrow \cdots\rightarrow v_i\rightarrow
Q\rightarrow v_j\cdots\rightarrow v_{t-1}\rangle$, where
$Q=\langle v_{i+1}\rightarrow v_{i+2}\rightarrow \cdots\rightarrow
v_{j-1} \rangle$. A path (or cycle) in $G$ is called a
\textit{Hamiltonian path} (or \textit{Hamiltonian cycle}) if it
contains every node of $G$ exactly once. Two paths (or cycles)
$P_1$ and $P_2$ connecting a node $u$ to a node $v$ are said to be
\textit{edge-disjoint} iff $E(P_1)\cap E(P_2)=\emptyset$. Two
paths (or cycles) $Q_1$ and $Q_2$ of graph $G$ are called
\textit{node-disjoint} iff $V(Q_1)\cap V(Q_2)=\emptyset$. Two
node-disjoint paths (or cycles) $Q_1$ and $Q_2$ of graph $G$ are
said to be \textit{equal} iff $|V(Q_1)|=|V(Q_2)|$ and $V(Q_1)\cup
V(Q_2)=V(G)$. Two node-disjoint paths $Q_1$ and $Q_2$ can be
\textit{concatenated} into a path, denoted by $Q_1\Rightarrow
Q_2$, if $end(Q_1)$ is adjacent to $start(Q_2)$.

Now, we introduce twisted cubes. The node set of the
$n$-dimensional twisted cube $TQ_n$ is the set of all binary
strings of length $n$, where $n$ is odd. A binary string $b$ of
length $n$ is denoted by $b_{n-1}b_{n-2}\cdots b_1b_0$, where
$b_{n-1}$ is the most significant bit. We denote the complement of
$b_i$ by $\overline{b}_i = 1-b_i$. To define $TQ_n$, a $i$-th bit
\textit{parity function} $\mathcal{P}_i(b)$ is introduced. Let
$b=b_{n-1}b_{n-2}\cdots b_1b_0$ be a binary string. For
$0\leqslant i\leqslant n-1$, $\mathcal{P}_i(b)=b_i\oplus
b_{i-1}\oplus\cdots\oplus b_1\oplus b_0$, where $\oplus$ is the
exclusive-or operation. Note that $0\oplus 0 = 1\oplus 1 = 0$ and
$0\oplus 1 = 1\oplus 0 = 1$. We then give the recursive definition
of the $n$-dimensional twisted cube $TQ_n$ for any odd integer
$n\geqslant 1$ as follows.

\begin{defn}\cite{Hilbers87,Yang09}\label{def_twisted-cube}
$TQ_1$ is the complete graph with two nodes labeled by 0 and 1,
respectively. For an odd integer $n \geqslant 3$, $TQ_n$ consists
of four copies of $TQ_{n-2}$. We use $TQ_{n-2}^{ij}$ to denote an
$(n-2)$-dimensional twisted cube which is a subgraph of $TQ_n$
induced by the nodes labeled by $ijb_{n-3}\cdots b_1b_0$, where
$i,j\in\{0,1\}$. Edges that connect these four subtwisted cubes
can be described as follows: Each node $b = b_{n-1}b_{n-2}\cdots
b_1b_0 \in V(TQ_n)$ is adjacent to
$\overline{b}_{n-1}b_{n-2}\cdots b_1b_0$ and
$\overline{b}_{n-1}\overline{b}_{n-2}\cdots b_1b_0$ if
$\mathcal{P}_{n-3}(b) = 0$; and to
$\overline{b}_{n-1}b_{n-2}\cdots b_1b_0$ and
$b_{n-1}\overline{b}_{n-2}\cdots b_1b_0$ if $\mathcal{P}_{n-3}(b)
= 1$.
\end{defn}

\begin{figure}[t]
\begin{center}
\includegraphics[scale=0.7]{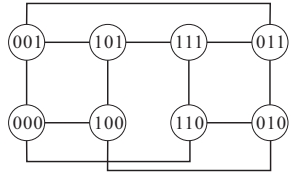}
\caption{The 3-dimensional twisted cube $TQ_3$.} \label{Fig_TQ_3}
\end{center}
\end{figure}

\begin{figure}[t]
\begin{center}
\includegraphics[scale=0.8]{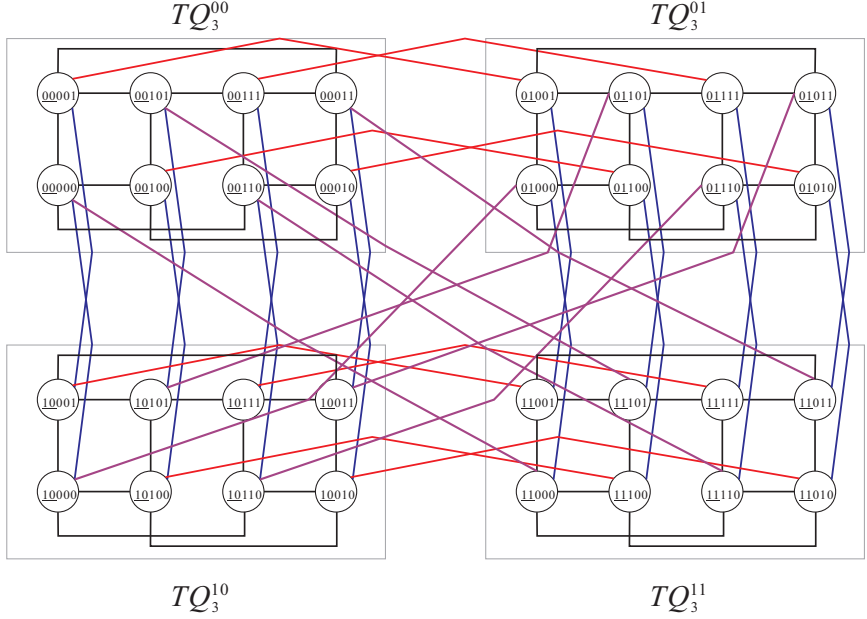}
\caption{The 5-dimensional twisted cube $TQ_5$ containing
$TQ_3^{00}$, $TQ_3^{10}$, $TQ_3^{01}$, $TQ_3^{11}$.}
\label{Fig_TQ_5}
\end{center}
\end{figure}

According to Definition \ref{def_twisted-cube}, $TQ_n$ is an
$n$-regular graph with $2^n$ nodes and $n2^{n-1}$ edges. The
parameter $n$ is always an odd integer if it is a dimension of the
twisted cube. In addition, $TQ_n$ is decomposed into four
subtwisted cubes $TQ_{n-2}^{00}$, $TQ_{n-2}^{10}$,
$TQ_{n-2}^{01}$, $TQ_{n-2}^{11}$, where $TQ_{n-2}^{ij}$ consists
of those nodes $b$ with $b_{n-1}=i$ and $b_{n-2}=j$. For each
$ij\in\{00,10,01,11\}$, $TQ_{n-2}^{ij}$ is isomorphic to
$TQ_{n-2}$. For example, Fig. \ref{Fig_TQ_3} shows $TQ_3$ and Fig.
\ref{Fig_TQ_5} depicts $TQ_5$ containing four subtwisted cubes
$TQ_3^{00}$, $TQ_3^{10}$, $TQ_3^{01}$, $TQ_3^{11}$.

Let $b$ is a binary string $b_{t-1}b_{t-2}\cdots b_1b_0$ of length
$t$. We denote $b^i$ the new binary string obtained by repeating
$b$ string $i$ times. Then, the length of $b^i$ is $i*t$. For
instance, $(01)^3=010101$ and $0^4=0000$.

\section{Two Edge-Disjoint Hamiltonian Cycles}\label{EdgeDisjointHC}
Obviously, $TQ_3$ has no two edge-disjoint Hamiltonian cycles
since each node is incident to three edges. Our method for
constructing two edge-disjoint Hamiltonian cycles of $TQ_n$, with
odd integer $n\geqslant 5$, is based on an inductive construction.
We will construct two edge-disjoint Hamiltonian paths $P$ and $Q$
in $TQ_n$, with $n\geqslant 5$, such that
$start(P)=00(0)^{n-5}000$, $end(P)=11(0)^{n-5}000$,
$start(Q)=00(0)^{n-5}100$, and $end(Q)=01(0)^{n-5}100$. The basic
idea is described as follows. Initially, we construct two
edge-disjoint Hamiltonian paths $P$ and $Q$ in $TQ_5$ such that
$start(P)=00000$, $end(P)=11000$, $start(Q)=00100$, and
$end(Q)=01100$. By the definition of parity function
$\mathcal{P}_i(\cdot)$, $\mathcal{P}_2(11000)=0$ and
$\mathcal{P}_2(01100)=1$. By Definition \ref{def_twisted-cube},
$start(P)\in N(end(P))$ and $start(Q)\in N(end(Q))$. Thus, $P$ and
$Q$ are two edge-disjoint Hamiltonian cycles. Consider that $n$ is
an odd integer with $n\geqslant 7$. We first partition $TQ_n$ into
four subtwisted cubes $TQ_{n-2}^{00}$, $TQ_{n-2}^{10}$,
$TQ_{n-2}^{01}$, $TQ_{n-2}^{11}$. Assume that $P^{ij}$ and
$Q^{ij}$ are two edge-disjoint Hamiltonian paths in
$TQ_{n-2}^{ij}$, for $i,j\in\{0,1\}$, such that
$start(P^{ij})=ij00(0)^{n-7}000$, $end(P^{ij})=ij11(0)^{n-7}000$,
$start(Q^{ij})=ij00(0)^{n-7}100$, and
$end(Q^{ij})=ij01(0)^{n-7}100$. We then discover six distinct
edges to concatenate these eight edge-disjoint paths into two
edge-disjoint Hamiltonian paths $P$ and $Q$ of $TQ_n$ such that
$start(P)=00(0)^{n-5}000$, $end(P)=11(0)^{n-5}000$,
$start(Q)=00(0)^{n-5}100$, and $end(Q)=01(0)^{n-5}100$. By
Definition \ref{def_twisted-cube}, $P$ and $Q$ are two
edge-disjoint Hamiltonian cycles in $TQ_n$ since
$start(P)=00(0)^{n-5}000\in N(end(P))$ and
$start(Q)=00(0)^{n-5}100\in N(end(Q))$. The concatenating process
will be shown in Lemma \ref{2HP-TQ}.

Now, we first show that $TQ_5$ contains two edge-disjoint
Hamiltonian paths in the following lemma.

\begin{figure}[t]
\begin{center}
\includegraphics[scale=0.8]{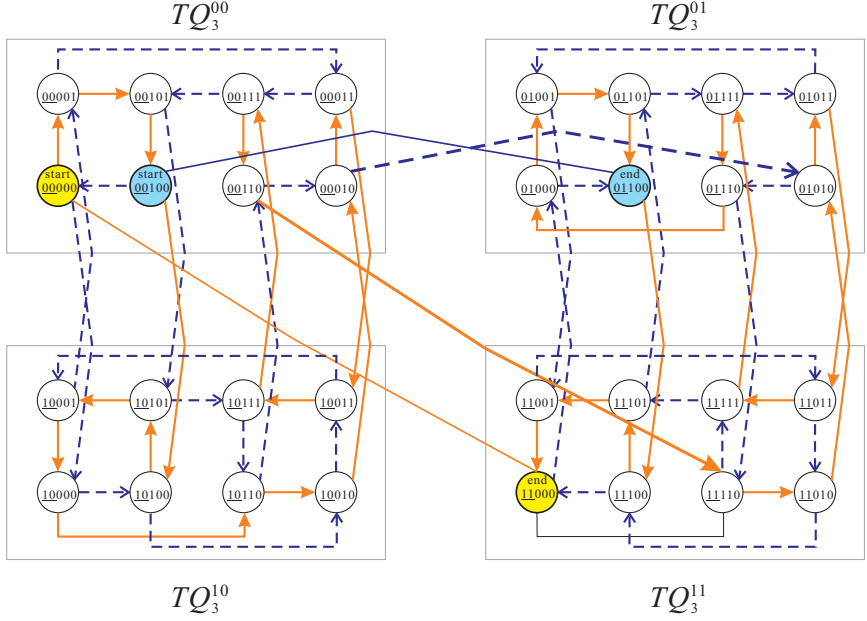}
\caption{Two edge-disjoint Hamiltonian paths in $TQ_5$, where
solid arrow lines indicate a Hamiltonian path and dotted arrow
lines indicate the other edge-disjoint Hamiltonian path.}
\label{Fig_TQ_5-2HP}
\end{center}
\end{figure}

\begin{lem}\label{2HP-TQ_5}
There are two edge-disjoint Hamiltonian paths $P$ and $Q$ in
$TQ_5$ such that $start(P)=00000$, $end(P)=11000$,
$start(Q)=00100$, and $end(Q)=01100$.
\end{lem}
\begin{proof}
We prove this lemma by constructing such two paths. Let\\
$P=\langle$00000 $\rightarrow$ 00001 $\rightarrow$ 00101
$\rightarrow$ 00100 $\rightarrow$ 10100 $\rightarrow$ 10101
$\rightarrow$ 10001 $\rightarrow$ 10000 $\rightarrow$ 10110
$\rightarrow$ 10010 $\rightarrow$ 00010 $\rightarrow$ 00011
$\rightarrow$ 10011 $\rightarrow$ 10111 $\rightarrow$ 00111
$\rightarrow$ 00110 $\rightarrow$ 11110 $\rightarrow$ 11010
$\rightarrow$ 01010 $\rightarrow$ 01011 $\rightarrow$ 11011
$\rightarrow$ 11111 $\rightarrow $01111 $\rightarrow$ 01110
$\rightarrow$ 01000 $\rightarrow$ 01001 $\rightarrow$ 01101
$\rightarrow$ 01100 $\rightarrow$ 11100 $\rightarrow$ 11101
$\rightarrow$ 11001 $\rightarrow$ 11000$\rangle$, and let\\
$Q=\langle$00100 $\rightarrow$ 00000 $\rightarrow$ 10000
$\rightarrow$ 10100 $\rightarrow$ 10010 $\rightarrow$ 10011
$\rightarrow$ 10001 $\rightarrow$ 00001 $\rightarrow$ 00011
$\rightarrow$ 00111 $\rightarrow$ 00101 $\rightarrow$ 10101
$\rightarrow$ 10111 $\rightarrow$ 10110 $\rightarrow$ 00110
$\rightarrow$ 00010 $\rightarrow$ 01010 $\rightarrow$ 01110
$\rightarrow$ 11110 $\rightarrow$ 11111 $\rightarrow$ 11101
$\rightarrow$ 01101 $\rightarrow$ 01111 $\rightarrow$ 01011
$\rightarrow$ 01001 $\rightarrow$ 11001 $\rightarrow$ 11011
$\rightarrow$ 11010 $\rightarrow$ 11100 $\rightarrow$ 11000
$\rightarrow$ 01000 $\rightarrow$ 01100$\rangle$.\\
Fig. \ref{Fig_TQ_5-2HP} depicts the constructions of $P$ and $Q$.
Clearly, $P$ and $Q$ are edge-disjoint Hamiltonian paths in
$TQ_5$.
\end{proof}

By Definition \ref{def_twisted-cube}, nodes 00000 and 11000 are
adjacent, and nodes 00100 and 01100 are adjacent. The following
corollary immediately holds true from Lemma \ref{2HP-TQ_5}.

\begin{cor}\label{2HC-TQ_5}
There are two edge-disjoint Hamiltonian cycles in $TQ_5$.
\end{cor}

Using Lemma \ref{2HP-TQ_5}, we prove the following lemma.

\begin{figure}[t]
\begin{center}
\includegraphics[scale=0.8]{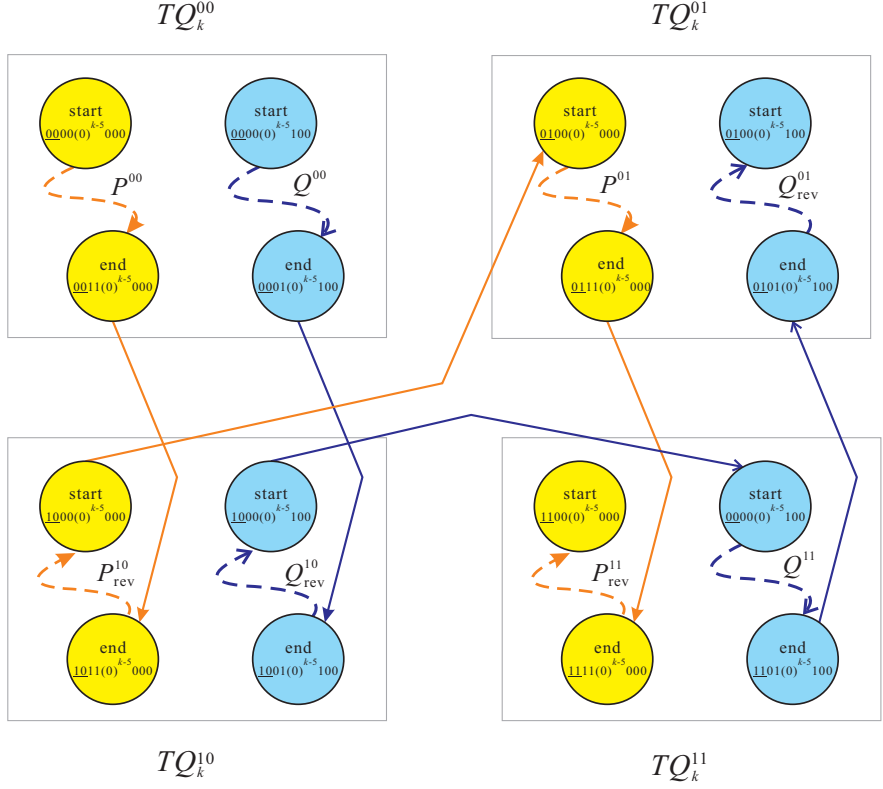}
\caption{The constructions of two edge-disjoint Hamiltonian paths
in $TQ_{k+2}$, with $k\geqslant 5$, where dotted arrow lines
indicate the paths and solid arrow lines indicate concatenated
edges.} \label{Fig_TQ_k-2HP}
\end{center}
\end{figure}

\begin{lem}\label{2HP-TQ}
For any odd integer $n\geqslant 5$, there are two edge-disjoint
Hamiltonian paths $P$ and $Q$ in $TQ_n$ such that
$start(P)=00(0)^{n-5}000$, $end(P)=11(0)^{n-5}000$,
$start(Q)=00(0)^{n-5}100$, and $end(Q)=01(0)^{n-5}100$.
\end{lem}
\begin{proof}
We prove this lemma by induction on $n$. By Lemma \ref{2HP-TQ_5},
the lemma holds true when $n=5$. Assume that the lemma holds when
$n=k\geqslant 5$. We will prove that the lemma holds true for
$n=k+2$. We first partition $TQ_{k+2}$ into four subtwisted cubes
$TQ_{k}^{00}$, $TQ_{k}^{10}$, $TQ_{k}^{01}$, $TQ_{k}^{11}$. By the
induction hypothesis, there are two edge-disjoint Hamiltonian
paths $P^{ij}$ and $Q^{ij}$, for $i,j\in\{0,1\}$, in $TQ_k^{ij}$
such that $start(P^{ij})=ij00(0)^{k-5}000$,
$end(P^{ij})=ij11(0)^{k-5}000$, $start(Q^{ij})=ij00(0)^{k-5}100$,
and $end(Q^{ij})=ij01(0)^{k-5}100$. By the definition of parity
function $\mathcal{P}_i(\cdot)$,
$\mathcal{P}_{k-1}(end(P^{ij}))=\mathcal{P}_{k-1}(start(P^{ij}))=0$,
$\mathcal{P}_{k-1}(end(Q^{ij}))=0$, and
$\mathcal{P}_{k-1}(start(Q^{ij}))=1$. According to Definition
\ref{def_twisted-cube}, we have that\\
$end(P^{00})\in N(end(P^{10}))$, $start(P^{10})\in
N(start(P^{01}))$, $end(P^{01})\in N(end(P^{11}))$,\\
$end(Q^{00})\in N(end(Q^{10}))$, $start(Q^{10})\in
N(start(Q^{11}))$, and $end(Q^{11})\in N(end(Q^{01}))$.\\
Let $P=P^{00} \Rightarrow P_{\textrm{rev}}^{10} \Rightarrow P^{01}
\Rightarrow P_{\textrm{rev}}^{11}$ and let $Q=Q^{00} \Rightarrow
Q_{\textrm{rev}}^{10} \Rightarrow Q^{11} \Rightarrow
Q_{\textrm{rev}}^{01}$, where $P_{\textrm{rev}}^{10}$,
$P_{\textrm{rev}}^{11}$, $Q_{\textrm{rev}}^{10}$, and
$Q_{\textrm{rev}}^{01}$ are the reversed paths of $P^{10}$,
$P^{11}$, $Q^{10}$, and $Q^{01}$, respectively. Then, $P$ and $Q$
are two edge-disjoint Hamiltonian paths in $TQ_{k+2}$ such that
$start(P)=00(0)^{k-3}000$, $end(P)=11(0)^{k-3}000$,
$start(Q)=00(0)^{k-3}100$, and $end(Q)=01(0)^{k-3}100$. Fig.
\ref{Fig_TQ_k-2HP} shows the constructions of such two
edge-disjoint Hamiltonian paths in $TQ_{k+2}$. Thus, the lemma
hods true when $n=k+2$. By induction, the lemma holds true.
\end{proof}

By Definition \ref{def_twisted-cube}, nodes
$start(P)=00(0)^{n-5}000$ and $end(P)=11(0)^{n-5}000$ are
adjacent, and nodes $start(Q)=00(0)^{n-5}100$ and
$end(Q)=01(0)^{n-5}100$ are adjacent. It immediately follows from
Lemma \ref{2HP-TQ} that the following corollary holds true.

\begin{cor}\label{2HC-TQ}
For any odd integer $n\geqslant 5$, there are two edge-disjoint
Hamiltonian cycles in $TQ_n$.
\end{cor}

\section{Two Equal Node-Disjoint Cycles}\label{NodeDisjointCyles}
In this section, we will construct two equal node-disjoint cycles
$P$ and $Q$ in a $n$-dimensional twisted cube $TQ_n$ with an odd
integer $n\geqslant 3$. Our method for constructing two equal
node-disjoint cycles of $TQ_n$ is also based on an inductive
construction. For any odd integer $n\geqslant 3$, we will
construct two equal node-disjoint paths $P$ and $Q$ in $TQ_n$ such
that $start(P)=00(0)^{n-3}1$, $end(P)=01(0)^{n-3}1$,
$start(Q)=00(0)^{n-3}0$, and $end(Q)=11(0)^{n-3}0$. The basic idea
is similar to that of constructing two edge-disjoint Hamiltonian
paths and is described as follows. Initially, we construct two
equal node-disjoint paths $P$ and $Q$ in $TQ_3$ such that
$start(P)=001$, $end(P)=011$, $start(Q)=000$, and $end(Q)=110$. By
Definition \ref{def_twisted-cube}, $P$ and $Q$ are also
node-disjoint cycles with the same length since their end nodes
are adjacent. Consider that $n$ is an odd integer with $n\geqslant
5$. We first partition $TQ_n$ into four subtwisted cubes
$TQ_{n-2}^{00}$, $TQ_{n-2}^{10}$, $TQ_{n-2}^{01}$,
$TQ_{n-2}^{11}$. Assume that $P^{ij}$ and $Q^{ij}$ are two equal
node-disjoint paths in $TQ_{n-2}^{ij}$, for $i,j\in\{0,1\}$, such
that $start(P^{ij})=ij00(0)^{n-5}1$, $end(P^{ij})=ij01(0)^{n-5}1$,
$start(Q^{ij})=ij00(0)^{n-5}0$, and $end(Q^{ij})=ij11(0)^{n-5}0$.
We then concatenate them into two equal node-disjoint paths $P$
and $Q$ of $TQ_n$ such that $start(P)=00(0)^{n-3}1$,
$end(P)=01(0)^{n-3}1$, $start(Q)=00(0)^{n-3}0$, and
$end(Q)=11(0)^{n-3}0$. By Definition \ref{def_twisted-cube}, $P$
and $Q$ are also two equal node-disjoint cycles of $TQ_n$ since
$start(P)\in N(end(P))$ and $start(Q)\in N(end(Q))$. The
concatenating process will be presented in Lemma \ref{2PC-TQ}.

For $TQ_3$, let $P=\langle$001 $\rightarrow$ 101 $\rightarrow$ 111
$\rightarrow$ 011$\rangle$ and let $Q=\langle$000 $\rightarrow$
100 $\rightarrow$ 010 $\rightarrow$ 110$\rangle$. Then, $P$ and
$Q$ are two equal node-disjoint paths in $TQ_3$. By Definition
\ref{def_twisted-cube}, $start(P)\in N(end(P))$ and $start(Q)\in
N(end(Q))$. Thus, the following lemma holds true.

\begin{lem}\label{2PC-TQ_3}
There are two equal node-disjoint paths $P$ and $Q$ in $TQ_3$ such
that $start(P)=001$, $end(P)=011$, $start(Q)=000$, and
$end(Q)=110$. Moreover, $P$ and $Q$ are two equal node-disjoint
cycles of $TQ_3$.
\end{lem}

\begin{lem}\label{2PC-TQ_3}
There are two equal node-disjoint paths $P$ and $Q$ in $TQ_3$ such
that $start(P)=001$, $end(P)=011$, $start(Q)=000$, and
$end(Q)=110$. Moreover, $P$ and $Q$ are two equal node-disjoint
cycles of $TQ_3$.
\end{lem}

Using Lemma \ref{2PC-TQ_3}, we prove the following lemma.

\begin{figure}[t]
\begin{center}
\includegraphics[scale=0.8]{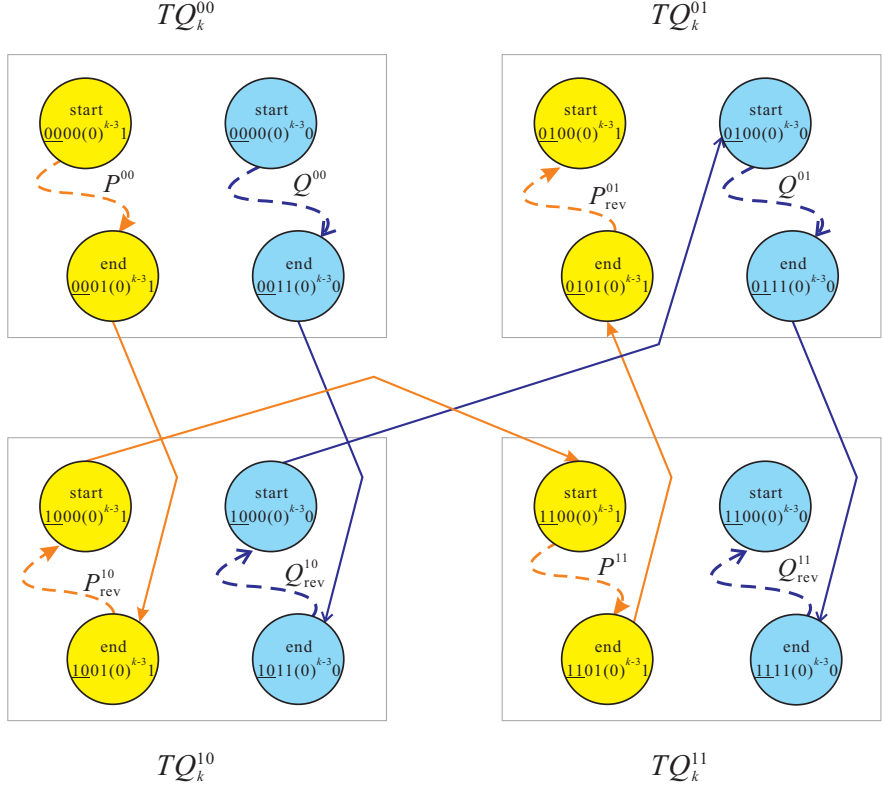}
\caption{The constructions of two equal node-disjoint paths in
$TQ_{k+2}$, with $k\geqslant 3$, where dotted arrow lines indicate
the paths and solid arrow lines indicate concatenated edges.}
\label{Fig_TQ_k-2PC}
\end{center}
\end{figure}

\begin{lem}\label{2PC-TQ}
For any odd integer $n\geqslant 3$, there are two equal
node-disjoint paths $P$ and $Q$ in $TQ_n$ such that
$start(P)=00(0)^{n-3}1$, $end(P)=01(0)^{n-3}1$,
$start(Q)=00(0)^{n-3}0$, and $end(Q)=11(0)^{n-3}0$.
\end{lem}
\begin{proof}
We prove this lemma by induction on $n$. By Lemma \ref{2PC-TQ_3},
the lemma holds true when $n=3$. Assume that the lemma holds when
$n=k\geqslant 3$. We will prove that the lemma holds true for
$n=k+2$. We first partition $TQ_{k+2}$ into four subtwisted cubes
$TQ_{k}^{00}$, $TQ_{k}^{10}$, $TQ_{k}^{01}$, $TQ_{k}^{11}$. By the
induction hypothesis, there are two equal node-disjoint paths
$P^{ij}$ and $Q^{ij}$, for $i,j\in\{0,1\}$, in $TQ_k^{ij}$ such
that $start(P^{ij})=ij00(0)^{k-3}1$, $end(P^{ij})=ij01(0)^{k-3}1$,
$start(Q^{ij})=ij00(0)^{k-3}0$, and $end(Q^{ij})=ij11(0)^{k-3}0$.
By the definition of parity function $\mathcal{P}_i(\cdot)$,
$\mathcal{P}_{k-1}(end(P^{ij}))=0$,
$\mathcal{P}_{k-1}(start(P^{ij}))=1$, and
$\mathcal{P}_{k-1}(end(Q^{ij}))=\mathcal{P}_{k-1}(start(Q^{ij}))=0$.
According to Definition \ref{def_twisted-cube}, we have that\\
$end(P^{00})\in N(end(P^{10}))$, $start(P^{10})\in
N(start(P^{11}))$, $end(P^{11})\in N(end(P^{01}))$,\\
$end(Q^{00})\in N(end(Q^{10}))$, $start(Q^{10})\in
N(start(Q^{01}))$, and $end(Q^{01})\in N(end(Q^{11}))$.\\
Let $P=P^{00} \Rightarrow P_{\textrm{rev}}^{10} \Rightarrow P^{11}
\Rightarrow P_{\textrm{rev}}^{01}$ and let $Q=Q^{00} \Rightarrow
Q_{\textrm{rev}}^{10} \Rightarrow Q^{01} \Rightarrow
Q_{\textrm{rev}}^{11}$, where $P_{\textrm{rev}}^{10}$,
$P_{\textrm{rev}}^{01}$, $Q_{\textrm{rev}}^{10}$, and
$Q_{\textrm{rev}}^{11}$ are the reversed paths of $P^{10}$,
$P^{01}$, $Q^{10}$, and $Q^{11}$, respectively. Then, $P$ and $Q$
are two equal node-disjoint paths in $TQ_{k+2}$ such that
$start(P)=00(0)^{k-1}1$, $end(P)=01(0)^{k-1}1$,
$start(Q)=00(0)^{k-1}0$, and $end(Q)=11(0)^{k-1}0$. Fig.
\ref{Fig_TQ_k-2PC} depicts the constructions of such two equal
node-disjoint paths $P$ and $Q$ in $TQ_{k+2}$. Thus, the lemma
hods true when $n=k+2$. By induction, the lemma holds true.
\end{proof}

By Definition \ref{def_twisted-cube}, nodes
$start(P)=00(0)^{n-3}1$ and $end(P)=01(0)^{n-3}1$ are adjacent,
and nodes $start(Q)=00(0)^{n-3}0$ and $end(Q)=11(0)^{n-3}0$ are
adjacent. It immediately follows from Lemma \ref{2PC-TQ} that the
following corollary holds true.

\begin{cor}\label{2CC-TQ}
For any odd integer $n\geqslant 3$, there are two equal
node-disjoint cycles in $TQ_n$.
\end{cor}

\section{Concluding Remarks}\label{Conclusion}
In this paper, we construct two edge-disjoint Hamiltonian cycles
(paths) of a $n$-dimensional twisted cubes $TQ_n$, for any odd
integer $n\geqslant 5$. On the other hand, we also construct two
equal node-disjoint cycles (paths) of $TQ_n$, for any odd integer
$n\geqslant 3$. In the construction of two edge-disjoint
Hamiltonian cycles (paths) of $TQ_n$, some edges are not used. It
is interesting to see if there are more edge-disjoint Hamiltonian
cycles of $TQ_n$ for $n\geqslant 7$. We would like to post it as
an open problem to interested readers.



\begin{thebibliography}{99}

\bibitem{Abraham91}
S. Abraham and K. Padmanabhan, The twisted cube topology for
multiprocessors: A study in network asymmetry, J. Parallel
Distrib. Comput. 13 (1991) 104--110.

\bibitem{Bae03}
M.M. Bae and B. Bose, Edge disjoint Hamiltonian cycles in $k$-ary
$n$-cubes and hypercubes, IEEE Trans. Comput. 52(10) (2003)
1271--1284.

\bibitem{Barden99}
B. Barden, R. Libeskind-Hadas, J. Davis, and W. Williams, On
edge-disjoint spanning trees in hypercubes, Inform. Process. Lett.
70 (1999) 13--16

\bibitem{Bhuyan84}
L.N. Bhuyan and D.P. Agrawal, Generalized hypercube and hyperbus
structures for a computer network, IEEE Trans. Comput. C-33(4)
(1984) 323--333.

\bibitem{Chang99}
C.P. Chang, J.N. Wang, and L.H. Hsu, Topological properties of
twisted cubes, Inform. Sci. 113 (1999) 147--167.

\bibitem{Choudum02}
S.A. Choudum and V. Sunitha, Augmented cubes, Networks 40(2)
(2002) 71--84.

\bibitem{Cull95}
P. Cull and S.M. Larson, The M\"{o}bius cubes, IEEE Trans. Comput.
44(5) (1995) 647--659.

\bibitem{Du88}
D.Z. Du and F.K. Hwang, Generalized de Bruijn digraphs, Networks
18 (1988) 27--38.

\bibitem{Efe92}
K. Efe, The crossed cube architecture for parallel computing, IEEE
Trans. Parallel Distribut. Syst. 3(5) (1992) 513--524.

\bibitem{Fan08}
J. Fan, X. Jia, and X. Lin, Embedding of cycles in twisted cubes
with edge pancyclic, Algorithmica 51 (2008) 264--282.

\bibitem{Fu08}
J.S. Fu, Fault-free Hamiltonian cycles in twisted cubes with
conditional link faults, Theoret. Comput. Sci. 407 (2008)
318--329.

\bibitem{Hilbers87}
P.A.J. Hilbers, M.R.J. Koopman, and J.L.A. van de Snepscheut, The
twisted cube, in: Lecture Notes in Comput. Sci., Parallel
Architect. Lang. Eur. (1987) 152--159.

\bibitem{Hsieh09}
S.Y. Hsieh and C.J. Tu, Constructing edge-disjoint spanning trees
in locally twisted cubes, Theoret. Comput. Sci. 410 (2009)
926--932.

\bibitem{Hsieh10}
S.Y. Hsieh and C.Y. Wu, Edge-fault-tolerant Hamiltonicity of
locally twisted cubes under conditional edge faults, J. Comb.
Optim. 19 (2010) 16--30.

\bibitem{Huang02}
W.T. Huang, J.M. Tan, C.N. Hung, and L.H. Hsu, Fault-tolerant
Hamiltonianicity of twisted cubes, J. Parallel Distrib. Comput. 62
(2002) 591--604.

\bibitem{Hwang00}
S.C. Hwang and G.H. Chen, Cycles in butterfly graphs, Networks
35(2) (2000) 161--171.


\bibitem{Lai08}
C.J. Lai and C.H. Tsai, Embedding a family of meshes into twisted
cubes, Inform. Process. Lett. 108 (2008) 326--330.

\bibitem{Lee90}
S. Lee and K. G. Shin, Interleaved all-to-all reliable broadcast
on meshes and hypercubes, in: Proc. Int. Conf. Parallel
Processing, vol. 3, 1990, pp. 110--113.

\bibitem{Li06}
T.K. Li, M.C. Yang, J.M. Tan, and L.H. Hsu, On embedding cycle in
faulty twisted cubes, Inform. Sci. 176 (2006) 676--690.

\bibitem{Petrovic06}
V. Petrovic and C. Thomassen, Edge-disjoint Hamiltonian cycles in
hypertournaments, J. Graph Theory 51 (2006) 49--52.

\bibitem{Saad88}
Y. Saad and M.H. Schultz, Topological properties of hypercubes,
IEEE Trans. Comput. 37(7) (1988) 867--872.

\bibitem{Rowley91}
R. Rowley and B. Bose, Edge-disjoint Hamiltonian cycles in de
Bruijn networks, in: Proc. 6th Distributed Memory Computing
Conference, 1991, pp. 707--709.

\bibitem{Yang06}
M.C. Yang, T.K. Li, Jimmy J.M. Tan, and L.H. Hsu, On embedding
cycles into faulty twisted cubes, Inform. Sci. 176 (2006)
676--690.

\bibitem{Yang09}
M.C. Yang, Edge-fault-tolerant node-pancyclicity of twisted cubes,
Inform. Process. Lett. 109 (2009) 1206--1210.

\end{thebibliography}
\end{document}